\documentclass[usenatbib]{mn2e}
\usepackage{graphicx,tabularx,amsmath}
\pdfoutput=1

\numberwithin{equation}{section}

\title[Statistical comparison of clouds and star clusters]{Statistical comparison of clouds and star clusters}
\author[O. Lomax, A. P. Whitworth and A. Cartwright]{O. Lomax, A. P. Whitworth and A. Cartwright\\
School of Physics and Astronomy, Cardiff University, Cardiff CF24 3AA}

\date{Submitted 2010 September 13}

\begin{document}

\pagerange{\pageref{firstpage}--\pageref{lastpage}} \pubyear{2010}

\maketitle

\label{firstpage}

\begin{abstract}
The extent to which the projected distribution of stars in a cluster is due to a large-scale radial gradient, and the extent to which it is due to fractal sub-structure, can be quantified -- statistically -- using the measure ${\cal Q} = \bar{m}/\bar{s}$. Here $\bar{m}$ is the normalized mean edge length of its minimum spanning tree (i.e. the shortest network of edges connecting all stars in the cluster) and $\bar{s}$ is the correlation length (i.e. the normalized mean separation between all pairs of stars).

We show how ${\cal Q}$ can be indirectly applied to grey-scale images by decomposing the image into a distribution of points from which $\bar{m}$ and $\bar{s}$ can be calculated. This provides a powerful technique for comparing the distribution of dense gas in a molecular cloud with the distribution of the stars that condense out of it. We illustrate the application of this technique by comparing ${\cal Q}$ values from simulated clouds and star clusters.
\end{abstract}

\section{Introduction}

The dimensionless measure ${\cal Q}$ has been shown to be a robust discriminator between clusters with a large-scale radial density gradient and clusters with small-scale subclustering \citep{qpaper2004,StarFormingClusters,NatureAndNurture}.  As it stands, the ${\cal Q}$ method can only be reliably applied to a collection of points, i.e. stars.  However, given that star clusters are often embedded in gas clouds, it would be useful if ${\cal Q}$ could be adapted for grey-scale images, such as sub-millimetre maps.

For a two-dimensional cluster of points, ${\cal Q}$ is equal to the normalized mean edge length of the minimum spanning tree (MST) $\bar{m}$ divided by the normalized correlation length $\bar{s}$.  The value of $\bar{s}$ is defined as the mean separation between all points divided by the radius of the cluster. The MST is the shortest network of edges needed to connect together all the points in the cluster.  The value of $\bar{m}$ is its mean edge length normalised by the inverse square-root of the mean cluster surface density.  Neither value by itself can distinguish between large-scale radial clustering and small-scale fractal sub-clustering; however, $\bar{m}$ varies more with sub-clustering than $\bar{s}$, and $\bar{s}$ varies more with radial clustering than $\bar{m}$.  Because of this, the ratio ${\cal Q} = \bar{m}/\bar{s}$ can distinguish between the two, with ${\cal Q} > 0.8$ for radially clustered distributions and ${\cal Q} < 0.8$ for fractally sub-clustered ones.

\citet{qpaper2006} have shown that the correlation length can be adapted for use on grey-scale images as the brightness of a pixel is analogous to the surface-density of a cluster.  However, a robust grey-scale equivalent of the MST method has yet to be found.  We present an alternative to directly analysing grey-scale images; this method involves decomposing the image into a collection of points, for which ${\cal Q}$ can then be calculated.

In Section \ref{method} we describe how a $\bar{m}$ and $\bar{s}$ can be calculated from grey-scale images. Section \ref{testdata} details the models of star clusters and cloud images from which we calculate ${\cal Q}$. In Section \ref{discussion} we present and discuss ${\cal Q}$ values from both artifical clusters and clouds, conclusions are presented in Section \ref{conclusions}.

\section{Methodology}
\label{method}

\begin{figure}
   \centering
   \includegraphics[width=\columnwidth]{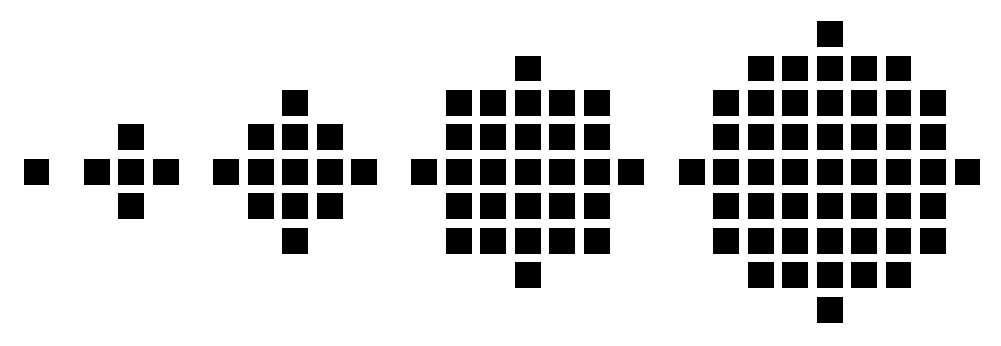}
   \caption{Starting from left to right, patches encompassing all pixels within $n=0,\,1,\,2,\,3\text{ and }4$ pixel-widths from a chosen pixel.}
   \label{cells}
\end{figure}

Suppose that we have a grey-scale image of $N_\text{pix}={\cal I}\times{\cal J}$ square pixels, each with the same angular size $\delta\times\delta$, and that the flux received from pixel $(i,j)$ is
\begin{equation}
   F_{ij}=\int_{\text{pixel}\,(i,j)}I\,d\Omega\,,
\end{equation}
where $I$ is the intensity (in whatever wavelength band is being used) and $d\Omega$ is an element of solid angle. It follows that the total flux received is
\begin{equation}
   F_\text{tot}=\sum_{i=1}^{i={\cal I}}\sum_{j=1}^{j={\cal J}}\{F_{ij}\}\,.
\end{equation}

We decompose the image into $N_\text{pnt}$ points. The choice of $N_\text{pnt}$ is discussed later. The flux quantum is then chosen to be
\begin{equation}
   \Delta F=\frac{F_\text{tot}}{N_\text{pnt}}\,.
\end{equation}
To convert the grey-scale image into an equivalent array of points, we pick a pixel ${\cal R}(i,j)$ at random, with no account taken of its flux $F_{ij}$.

If $F_{ij}\geq \Delta F$, then we reduce
\begin{equation}
   F_{ij}\to F_{ij}-\Delta F\,,
   \label{reduce1}
\end{equation}
and place a point at ${\bf r}_{ij}+\Delta{\bf r}_\text{rnd}$, where ${\bf r}_{ij}$ is the centre of pixel ${\cal R}(i,j)$ and $\Delta{\bf r}_\text{rnd}$ is a small random displacement (smaller in magnitude than the linear size of a pixel, $\delta$).

If $F_{ij}<\Delta F$, we consider a patch of pixels. An $n$-patch comprises all the pixels whose  centres lie at angular separation $\leq n\delta$ from the centre of pixel ${\cal R}(i,j)$; the configuration of $n$-patches for $n=0,\,1,\,2,\,3\text{ and }4$ are illustrated on Figure \ref{cells}. We increase $n$ until the flux from the $n$-patch exceeds or equals $\Delta F$, i.e.
\begin{equation}
   F_{n\text{-patch}}=\sum_{n\text{-patch}}\{F_{ij}\}\geq\Delta F\,.
\end{equation}
We then reduce the flux from each pixel within the $n$-patch, {\it pro-rata}, i.e.
\begin{equation}
   F_{ij}\to F_{ij}\left(1-\frac{\Delta F}{F_{n\text{-patch}}}\right)\,
   \label{reduce2}
\end{equation}
and place a point at position ${\bf r}_\text{pnt}$ which is equal to the weighted centre of the removed flux, plus a small random displacement, i.e.
\begin{equation}
   {\bf r}_\text{pnt}=\sum_{n\text{-patch}}\left\{\frac{F_{ij}\,{\bf r}_{ij}}{F_{n\text{-patch}}}\right\}+\Delta {\bf r}_\text{rnd}\,.
\end{equation}
We repeat this process $N_\text{pnt}$ times, thus reducing $F_\text{tot}$ to zero:
\begin{equation}
   F_\text{tot}\to F_\text{tot}-\sum_{i=1}^{i=N_\text{pnt}}\left\{ \frac{F_\text{tot}}{N_\text{pnt}}\right\}=0\,.
\end{equation}
Note that for every iteration of this algorithm, ${\cal R}(i,j)$ is chosen completely at random and is thus permitted to have the same value more than once. Also, flux ``detritus'' left over from equations (\ref{reduce1}) and (\ref{reduce2}) is invariably swept up and accounted for by later iterations; the final iteration, for example, has an $n$-patch size which encompasses the entire image.

From this collection of $N_\text{pnt}$ points, we can now generate the minimum spanning tree using Kruskal's algorithm \citep{kruskal} with normalized mean edge length
\begin{equation}
   \bar{m}=\frac{1}{\sqrt{A_N N}}\sum_{i=1}^{i=N-1}m_i\,,
   \label{mbar}
\end{equation}
where $N=N_\text{pnt}$, $A_N$ is the circular area of the point distribution (i.e. the smallest circle encompassing all points) and $m_i$ is the length of edge $i$ of the minimum spanning tree. The correlation length is then calculated
\begin{equation}
   \bar{s}=\frac{2}{N(N-1) R_N}\sum_{i=1}^{i=N-1}\sum_{j=i+1}^{j=N}{|{\bf r}_i-{\bf r}_j|}\,,
   \label{sbar}
\end{equation}
where $R_N=(A_N/\pi)^{1/2}$. Note that $\bar{s}$ is normalized by $R_N$ and as the number of edeges within a two-dimensional area scales as $N^2$. For $\bar{m}$, the number of edges scales as $N$, thus the average edge-length needs to be divided by $(N A)^{1/2}{(N-1)}$ (see \citet{qpaper2004} for more details).

It can be argued that a circular area is not always the most intuitive shape to consider for these purposes; for example, some work with ${\cal Q}$ specifically considers non-circular distributions \citep{qpaper_elong} and different definitions of the area, \citep{StarFormingClusters}. However, as $R_N=(A_N/\pi)^{1/2}$ in equation (\ref{sbar}), these area components cancel as ${\cal Q}=\bar{m}/\bar{s}\,$.

By picking the intial positions of sampling cells randomly (rather than, for example, starting with the brightest pixel, as is done in some clump-finding algorithms) and further adding $\Delta {\bf r}_\text{rnd}$ to ${\bf r}_\text{pnt}$, we help to break up the lattice structure native to a grey-scale image and ensure that no two points lie directly atop one another.

By averaging over patches of pixels, the algorithm smooths over small-scale intensity variations and hence lose information in converting images into points. These losses can be avoided by setting $N_\text{pnt}$ to a sufficiently high value. For a grey-scale image with discrete integer values assigned to each pixel, setting $N_\text{pnt}=F_\text{tot}$ would enable an exact reconstruction of the image from its sampled distribution of points. This would require approximately $N_\text{pnt}\sim{\cal I}\times{\cal J}\times 2^{N_\text{bpp}}$, where $N_\text{bpp}$ is the number of bits associated to the value of each pixel. For a $100\times100$ pixel image, this would correspond to $N_\text{pnt}\sim10^6$ for $N_\text{bpp}=8$ and $N_\text{pnt}\sim10^8$ for $N_\text{bpp}=16\,$. Whilst this is valid for images with discrete integer value pixels, astronomical images are often composed of continuous floating-point value pixels. In this case, a distribution of discrete points can not {\it exactly} represent a grey-scale image for any practical value of $N_\text{pnt}$.

Through preliminary testing, we find that for an image generated with a flux-distribution function $F(x,y)$, the resulting ${\cal Q}$ value is dependent on both image resolution and $N_\text{pnt}$, as shown in Figure \ref{resplot}. These dependencies can be largely mitigated by setting ratio of sampled points to pixels $N_\text{pnt}/N_\text{pix}$ to a constant value, again indicated in Figure \ref{resplot}. Unless otherwise stated, we have set $N_\text{pnt}=N_\text{pix}$ for the results presented in Section \ref{discussion}.

\section{Model clouds and clusters}
\label{testdata}
\subsection{Radial power-law distributions}
\label{rdist}

Centrally concentrated distributions of stars and interstellar gas can be constructed with with density
\begin{equation}
   \rho(r) = \rho_0\left(\frac{r}{r_0}\right)^{-\alpha}\,,
   \label{powerlaw}
\end{equation}
where $\rho(r)$ is the density at radius $r$, $\rho_0$ is a defined density at fixed radius $r_0$ and $\alpha$ is the density exponent. A synthetic star cluster is created randomly using the Monte-Carlo method to position stars according to equation (\ref{powerlaw}). Such a cluster contains $N_\star$ stars with positions
\begin{equation}
   \begin{split}
      r&={{\cal R}_r}^{1/(3-\alpha)}\,,\\
      \theta&=\cos^{-1}({2\cal R}_\theta-1)\,,\\
      \phi&=2\pi {\cal R}_\phi\,,\\
      x&=r\sin(\theta)\cos(\phi)\,,\\
      y&=r\sin(\theta)\sin(\phi)\,,\\
      z&=r\cos(\theta)\,,
   \end{split}
   \label{rcluster}
\end{equation}
where ${\cal R}_r,\,{\cal R}_\theta\text{ and }{\cal R}_\phi$ are random numbers between zero and one \citep{qpaper2004}.

For a gas cloud with a density profile given by equation (\ref{powerlaw}), the surface-density at impact parameter $b$ is
\begin{equation}
   \Sigma(b)=2b^{1-\alpha}\int_0^{\cos^{-1}(b)}\sec^{2-\alpha}(\theta)d\theta\,.
   \label{los1}
\end{equation}
This can be calculated analytically for integer $\alpha$:
\begin{equation}
   \Sigma(b)=\begin{cases}
      2(1-b^2)^{1/2} & \text{if } \alpha=0\,,\\
      2\ln{\frac{1+(1-b^2)^{1/2}}{b}} & \text{if } \alpha=1\,,\\
      2\frac{\cos^{-1}(b)}{b} & \text{if } \alpha=2\,.
    \end{cases}
    \label{los2}
\end{equation}
We can also solve equation (\ref{los1}) numerically for non-integer values of $\alpha$. Using equation (\ref{los2}), we can produce ${\cal I}\times{\cal J}$ pixel images by setting $b=0$ at the centre of the image and assigning each pixel a value of $\Sigma(b)$. Examples are shown in Figure \ref{montage}.

\subsection{Fractal distributions}
\label{fdist}

In contrast to centrally concentrated power-law clustering, multiscale sub-clustering in astronomy can be characterized using fractals. Fractals possess self-similar scaling defined by a fractal dimension $D$. Regular Euclidean geometry can also be shown to have a similar scaling property. For example, consider a cube of edge-length 1. If this is divided up into $N=8$ sub-cubes then each will have edge-length $l=1/\sqrt[3]{N}=1/2\,$. Assuming this relationship between $N$ and $l$ holds over all scales, it is said to have a fractal dimension of
\begin{equation}
   D=\frac{\log(N)}{\log(1/l)}=\frac{\log(8)}{\log(2)}=3\,.
\end{equation}
Now consider the same cube but this time populated with $N=4$ sub-cubes of edge-length $l=1/2$, i.e. four cubes and four cubic regions of empty space. If the sub-cubes are populated the same way over all scales, then it can be considered to have a fractal geometry with dimension $D=\log(4)/\log(2)=2\,$. In general, any three-dimensional shape that scales with $D<3$ is considered fractal \citep{FractalsInNature}.

A synthetic fractal star cluster of dimension $D$ can be constructed iteratively by considering a cube of edge-length 2 with a parent-star at its centre. The cube is then divided into eight sub-cubes, of which a random $2^D$ are given a child star at their centre. The parent star is then deleted with the children-stars becoming parents such that the process can be repeated over $N_\text{gen}$ generations. A little noise is added to the final positions of the stars to break the cubic structure and the distribution is pruned to a sphere of radius 1. This results in a fractally sub-clustered sphere of stars that is roughly self-similar down to a length-scale of $2\times2^{-N_\text{gen}}$. Fractals are often used in this way to generate clusters \citep{MeanSurfaceDensity,DynamicalEvolution,VeryWideBinaries}; conversely, as this paper details, there are also methods for extracting fractal information from real observations.

Following on from the synthesis of fractal star clusters, artificial fractal clouds of given $D$ can also be contructed. We start by considering a cube of edge-length 2 and uniform density 1. For generation $n=1$, the cube is split into 8 sub-cubes of $l=2\times2^{-n}=1$, of which a random $2^D$ mature and are given a density $\rho=8^n=8$ such that $\rho\,l^3=8\,$. The matured sub-cubes are recursively populated with sub-sub-cubes in the same way over $N_\text{gen}$ generations. In instances where $2^D$ is non-integer, the integer value is used to populate the current sub-cube and the remainder is passed on to the next sub-cube population. The process results in a cube divided into $8^{N_\text{gen}}$ sub-cubes of $l=2\times2^{-N_\text{gen}}$, approximating a fractal density field. The field is then pruned to a roughly spherical shape, i.e. $\rho=0$ for all sub-cubes outside a radius of 1. Note that throughout the recursive sub-division, the densities of unmatured sub-cubes are not set to zero. This serves to (i) introduce some residual background noise and (ii) avoid creating any ``vacuum'' within the density field.

\begin{figure}
   \includegraphics[width=\columnwidth]{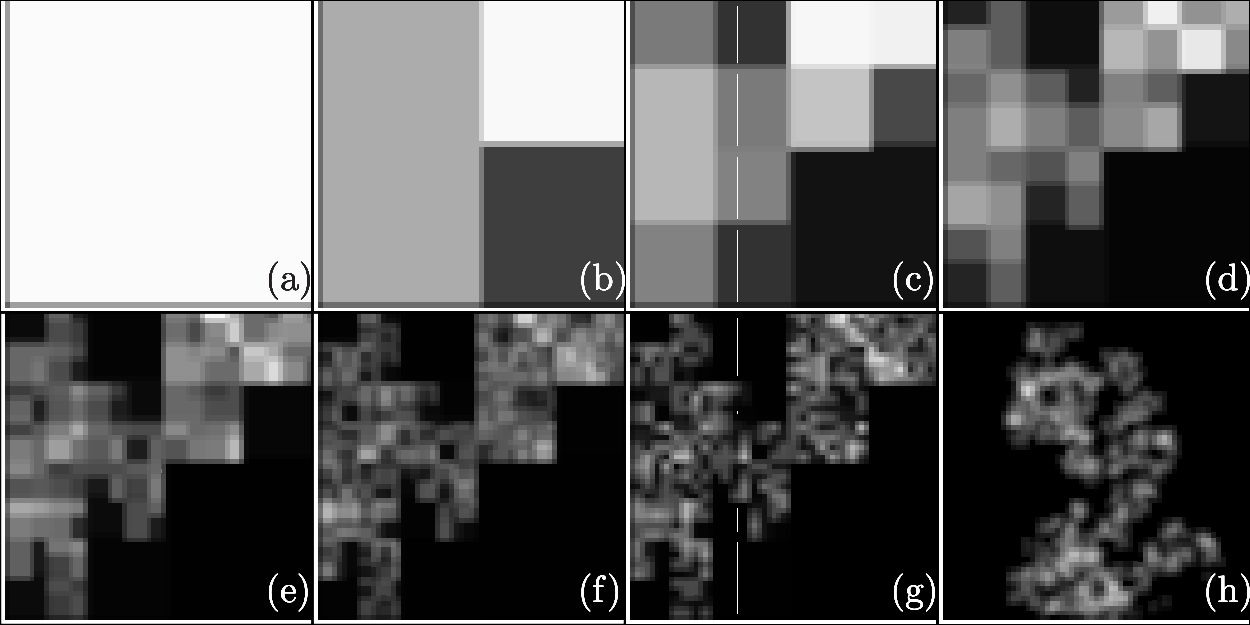}
   \caption{Generation of fractal cloud with $D=2.0\,$. Tiles show density fields projected onto the $x$-$y$ plane. Begining with a uniform medium in tile (a), we add fractal overdensities over six generations, as shown in tiles (b) to (g). Finallly, in tile (h) we crop the field into a sphere and project it on to an arbitrary plane.}
   \label{generations}
\end{figure}

We then project the column density of these clouds onto a 2-dimension plane through a random line of sight to create an image. Image resolution is chosen to reflect the scale of self-similarity for chosen $N_\text{gen}$, i.e. the image size is $2^{N_\text{gen}}\times2^{N_\text{gen}}$ pixels corresponding to $32\times32,\,64\times64,\,128\times128\text{ and }256\times256$ pixels for $N_\text{gen}=5,\,6,\,7\text{ and }8$ respectively. Figure \ref{generations} shows a step-by-step generation of a $D=2.0$, six-generation fractal cloud.

\subsection{Perimeter-area method}

One of the most common methods for measuring the three-dimensional fractal dimension $D$ from a grey-scale image is the perimeter-area method. This uses the relationship between the perimeter $P$ and area $A$ of a two-dimensional fractal shape
\begin{equation}
   P\propto A^{D_2/2}\,,
\end{equation}
where $D_2$ is the fractal dimension in two-dimensions \citep{FractalsInNature}. This can be applied to iso-contour lines from a grey-scale image, where a plot of $\log(P)$ against $\log(A)$ produces a slope of $D_2/2$. By measuring $D_2$ of known distributions, $D$ can be inferred (e.g. \citet{ProjectedClouds2005}). However, this method loses accuracy when $D>2.5$ and can not algorithmically distinguish between large-scale central clustering and multi-scale sub-clustering \citep{qpaper2006}.

\subsection{Acquiring ${\cal Q}$ statistics}
\label{acquiringQ}

\begin{figure*}
   \includegraphics[width=\textwidth]{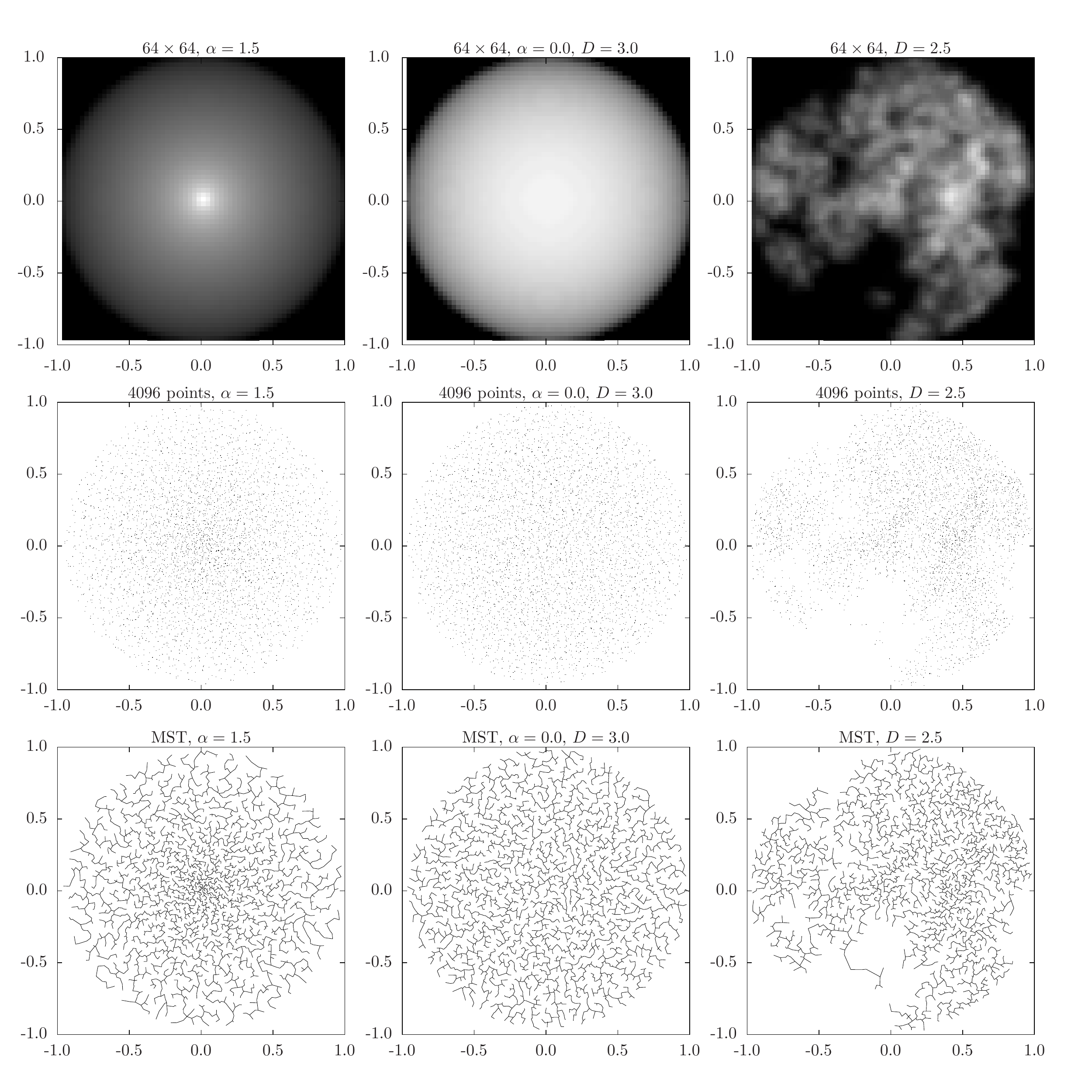}
   \caption{Point distributions and MSTs from radial and fractal grey-scale images. Row 1 shows the original grey-scale data, row 2 shows the distribution of points decomposed from the data and row 3 shows the MST connecting all the points. Column 1 has a radial density profile $\alpha=1.5\,$, column 2 is uniform density and column 3 has a fractal dimension $D=2.5$ .}
   \label{montage}
\end{figure*}

\begin{figure}
   \includegraphics[width=\columnwidth]{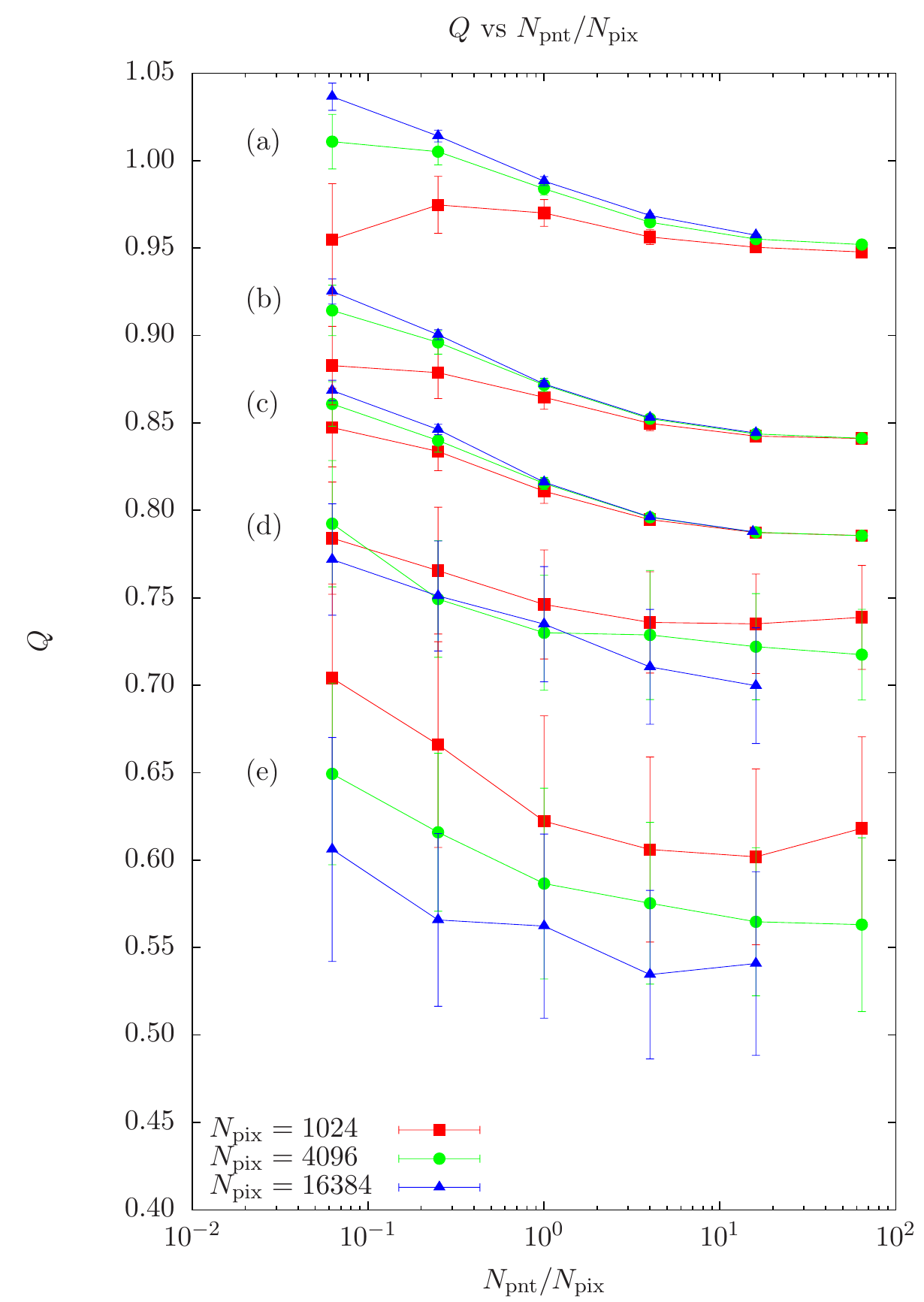}
   \caption{A plot of ${\cal Q}$ against the total number of points over the total number of pixels $N_\text{pnt}/N_\text{pix}$. The red squares represent ${\cal Q}$ values for $32\times32$ pixel images, the green circles for $64\times64$ pixel images and the blue triangles for $128\times128$ pixel images. The group of lines labeled (a) represent ${\cal Q}$ values for $\alpha=2.0\,$, (b) for $\alpha=1.0\,$, (c) for $\alpha=0.0$ and $D=3.0\,$, (d) for $D=2.5$ and (e) for $D=2.0\,$. Each data point is generated from fifty realisations.}
   \label{resplot}
\end{figure}

We generate radial grey-scale images with $2.0\ge \alpha \ge0.0$ and fractal grey-scale images with $3.0\ge D\ge2.0 $ for image sizes of $32\times32,\,64\times64,\,128\times128,\text{ and }256\times256$ pixels. Figure \ref{resplot} illustrates how the dependency of ${\cal Q}$ on image size is minimized when we set $N_\text{pnt}/N_\text{pix}=k$ where $k$ is a constant. For simplicity and computational manageability we use $k=1$, as discussed in Section \ref{method}.

The ${\cal Q}$ statistics from these images are shown in Table \ref{r_data} and compared with those from artificial clusters. A hundred realisations are performed for five values of $\alpha$ and five values of $D$ at each of the aforementioned image sizes. Examples of the grey-scale images, along with their sampled point distirbutions and minimum spanning trees, can be seen in Figure \ref{montage}.

The artificial star cluster ${\cal Q}$ statistics are generated from clusters of one hundred to one thousand points with specific radial and fractal distributions. As with the grey-scale images, one hundred realisations are performed for each value of $\alpha$ and $D$.

\section{Discussion}
\label{discussion}

\begin{table}
   \caption{Clustering statistics for radial-profile distributions. Note that $N=N_\text{pix}=N_\text{pnt}$ for all grey-scale images. Rows where $N=\text{``Cluster''}$ pertain to artificial star clusters with one hundred to one thousand stars. Statistical uncertainties with apparant zero values are at least an order of magnitude smaller than those with finite values.}
   \newcolumntype{R}{>{\raggedleft\arraybackslash}X}
   \newcolumntype{L}{>{\raggedright\arraybackslash}X}
   \newcolumntype{C}{>{\centering\arraybackslash}X}
   \begin{tabularx}{\columnwidth}{@{} c c C C C @{}}
      \hline
      $\alpha$ & $N$     & $\bar{m}$       & $\bar{s}$       & ${\cal Q}$\\
      \\
      \hline
      2.00     & 1024    & $0.572\pm0.006$ & $0.590\pm0.001$ & $0.969\pm0.009$ \\
               & 4096    & $0.576\pm0.003$ & $0.586\pm0.000$ & $0.984\pm0.005$ \\
               & 16384   & $0.577\pm0.001$ & $0.584\pm0.000$ & $0.989\pm0.002$ \\
               & 65536   & $0.575\pm0.001$ & $0.583\pm0.000$ & $0.988\pm0.001$ \\
               & Cluster & $0.544\pm0.012$ & $0.583\pm0.012$ & $0.932\pm0.020$ \\
      \hline
      1.50     & 1024    & $0.611\pm0.005$ & $0.673\pm0.001$ & $0.909\pm0.008$ \\
               & 4096    & $0.616\pm0.002$ & $0.673\pm0.000$ & $0.916\pm0.003$ \\
               & 16384   & $0.618\pm0.001$ & $0.673\pm0.000$ & $0.918\pm0.002$ \\
               & 65536   & $0.616\pm0.001$ & $0.673\pm0.000$ & $0.916\pm0.001$ \\
               & Cluster & $0.589\pm0.012$ & $0.674\pm0.008$ & $0.873\pm0.016$ \\
      \hline
      1.00     & 1024    & $0.634\pm0.005$ & $0.732\pm0.001$ & $0.866\pm0.007$ \\
               & 4096    & $0.638\pm0.002$ & $0.732\pm0.000$ & $0.871\pm0.003$ \\
               & 16384   & $0.639\pm0.001$ & $0.733\pm0.000$ & $0.873\pm0.002$ \\
               & 65536   & $0.638\pm0.001$ & $0.733\pm0.000$ & $0.871\pm0.001$ \\
               & Cluster & $0.613\pm0.011$ & $0.735\pm0.008$ & $0.833\pm0.015$ \\
      \hline
      0.50     & 1024    & $0.646\pm0.006$ & $0.774\pm0.001$ & $0.834\pm0.008$ \\
               & 4096    & $0.651\pm0.003$ & $0.775\pm0.000$ & $0.840\pm0.004$ \\
               & 16384   & $0.652\pm0.001$ & $0.776\pm0.000$ & $0.841\pm0.002$ \\
               & 65536   & $0.650\pm0.001$ & $0.776\pm0.000$ & $0.838\pm0.001$ \\
               & Cluster & $0.623\pm0.010$ & $0.775\pm0.011$ & $0.804\pm0.016$ \\
      \hline
      0.00     & 1024    & $0.653\pm0.005$ & $0.807\pm0.001$ & $0.809\pm0.006$ \\
               & 4096    & $0.658\pm0.002$ & $0.807\pm0.000$ & $0.815\pm0.003$ \\
               & 16384   & $0.659\pm0.001$ & $0.808\pm0.000$ & $0.816\pm0.002$ \\
               & 65536   & $0.657\pm0.001$ & $0.808\pm0.000$ & $0.815\pm0.001$ \\
               & Cluster & $0.633\pm0.011$ & $0.808\pm0.008$ & $0.783\pm0.014$ \\
      \hline
   \end{tabularx}
   \label{r_data}
\end{table}

\begin{table}
   \caption{Clustering statistics for fractal distributions.}
   \newcolumntype{R}{>{\raggedleft\arraybackslash}X}
   \newcolumntype{L}{>{\raggedright\arraybackslash}X}
   \newcolumntype{C}{>{\centering\arraybackslash}X}
   \begin{tabularx}{\columnwidth}{@{} c c C C C @{}}
      \hline
      $D$      & $N$     & $\bar{m}$       & $\bar{s}$       & ${\cal Q}$\\
      \\
      \hline
      3.00     & 1024    & $0.658\pm0.005$ & $0.809\pm0.001$ & $0.814\pm0.006$ \\
               & 4096    & $0.660\pm0.003$ & $0.808\pm0.000$ & $0.817\pm0.003$ \\
               & 16384   & $0.660\pm0.001$ & $0.808\pm0.000$ & $0.817\pm0.002$ \\
               & 65536   & $0.658\pm0.001$ & $0.808\pm0.000$ & $0.815\pm0.001$ \\
               & Cluster & $0.648\pm0.010$ & $0.809\pm0.010$ & $0.801\pm0.011$ \\
      \hline
      2.75     & 1024    & $0.615\pm0.020$ & $0.784\pm0.034$ & $0.785\pm0.023$ \\
               & 4096    & $0.618\pm0.019$ & $0.787\pm0.034$ & $0.785\pm0.021$ \\
               & 16384   & $0.612\pm0.021$ & $0.783\pm0.034$ & $0.782\pm0.022$ \\
               & 65536   & $0.616\pm0.022$ & $0.792\pm0.039$ & $0.778\pm0.021$ \\
               & Cluster & $0.609\pm0.022$ & $0.783\pm0.036$ & $0.779\pm0.025$ \\
      \hline
      2.50     & 1024    & $0.574\pm0.025$ & $0.769\pm0.047$ & $0.748\pm0.032$ \\
               & 4096    & $0.564\pm0.028$ & $0.772\pm0.054$ & $0.732\pm0.039$ \\
               & 16384   & $0.553\pm0.024$ & $0.764\pm0.050$ & $0.725\pm0.032$ \\
               & 65536   & $0.551\pm0.024$ & $0.767\pm0.058$ & $0.720\pm0.038$ \\
               & Cluster & $0.570\pm0.030$ & $0.777\pm0.049$ & $0.735\pm0.043$ \\
      \hline
      2.25     & 1024    & $0.516\pm0.030$ & $0.746\pm0.072$ & $0.695\pm0.045$ \\
               & 4096    & $0.495\pm0.027$ & $0.744\pm0.063$ & $0.669\pm0.045$ \\
               & 16384   & $0.471\pm0.029$ & $0.740\pm0.080$ & $0.641\pm0.052$ \\
               & 65536   & $0.464\pm0.022$ & $0.751\pm0.068$ & $0.622\pm0.047$ \\
               & Cluster & $0.494\pm0.030$ & $0.737\pm0.063$ & $0.673\pm0.052$ \\
      \hline
      2.00     & 1024    & $0.478\pm0.029$ & $0.739\pm0.079$ & $0.651\pm0.051$ \\
               & 4096    & $0.443\pm0.027$ & $0.742\pm0.080$ & $0.601\pm0.049$ \\
               & 16384   & $0.416\pm0.024$ & $0.737\pm0.073$ & $0.569\pm0.052$ \\
               & 65536   & $0.394\pm0.023$ & $0.744\pm0.076$ & $0.535\pm0.054$ \\
               & Cluster & $0.453\pm0.030$ & $0.739\pm0.084$ & $0.619\pm0.067$ \\
      \hline
   \end{tabularx}
   \label{f_data}
\end{table}

\begin{figure}
   \includegraphics[width=\columnwidth]{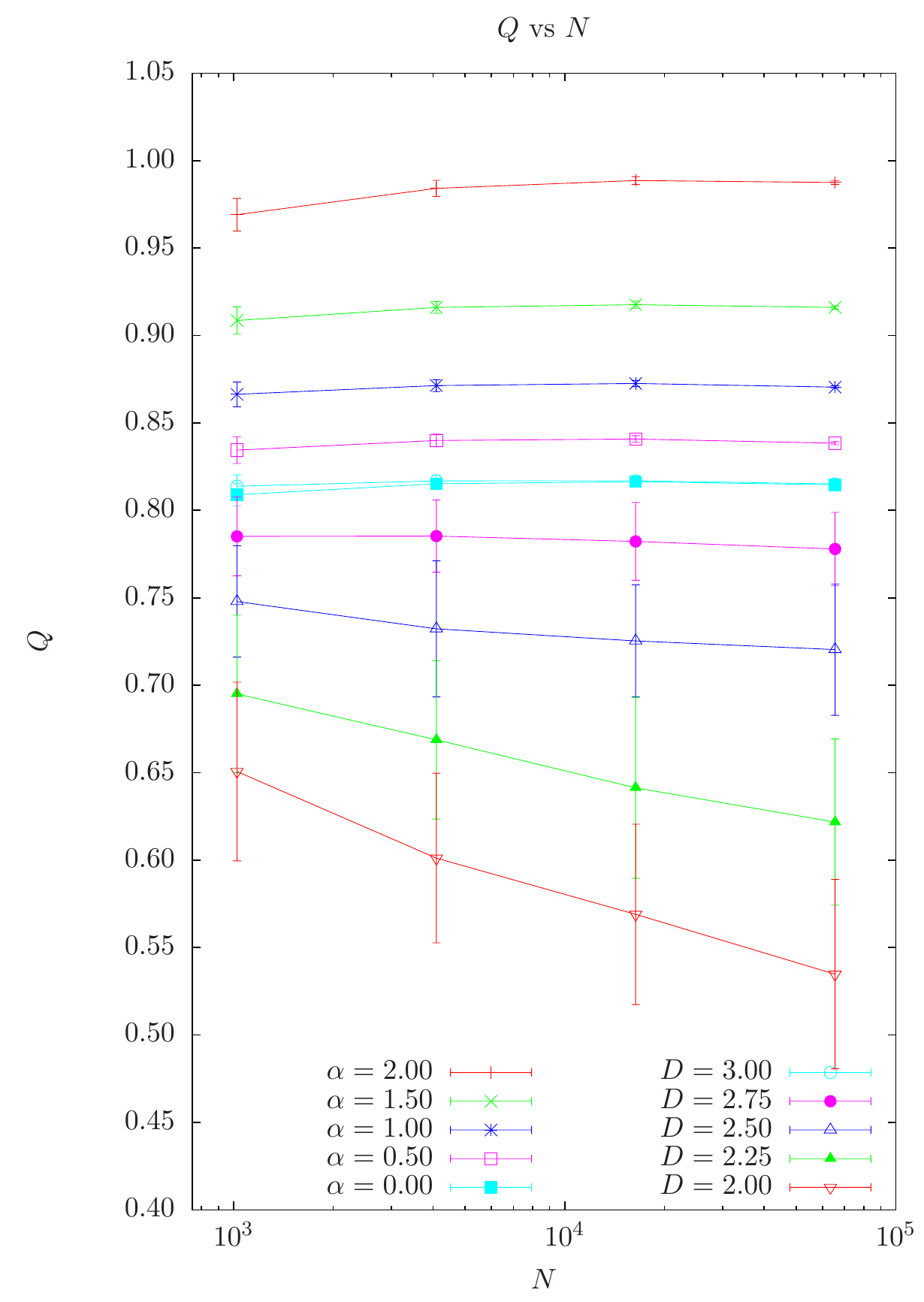}
   \caption{A plot of ${\cal Q}$ against the total number of pixels $N$. Different data series represent different values of $\alpha$ and $D$. Note that the statistical uncertainties on the radial ${\cal Q}$ values compared with the fractal ${\cal Q}$ values are negligable due to the isotropy of radial profile distributions.}
   \label{q_values}
\end{figure}

\begin{figure}
   \centering
   \includegraphics[width=\columnwidth]{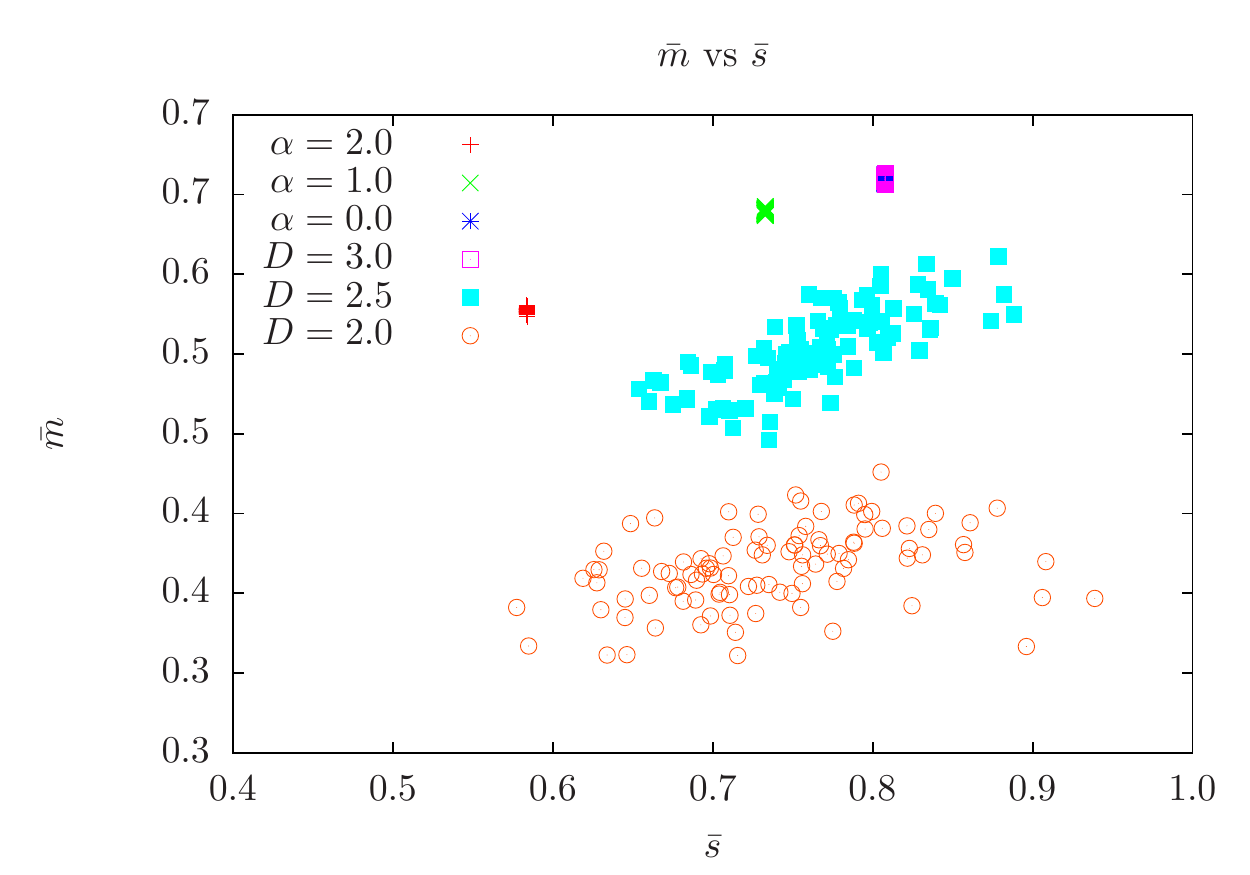}
   \caption{A plot of $\bar{m}$ against $\bar{s}$ for artificial cloud data. Each data set represents one hundred realisations of $128\times128$ pixel images.}
   \label{msbar}
\end{figure}

\begin{figure}
   \centering
   \includegraphics[width=\columnwidth]{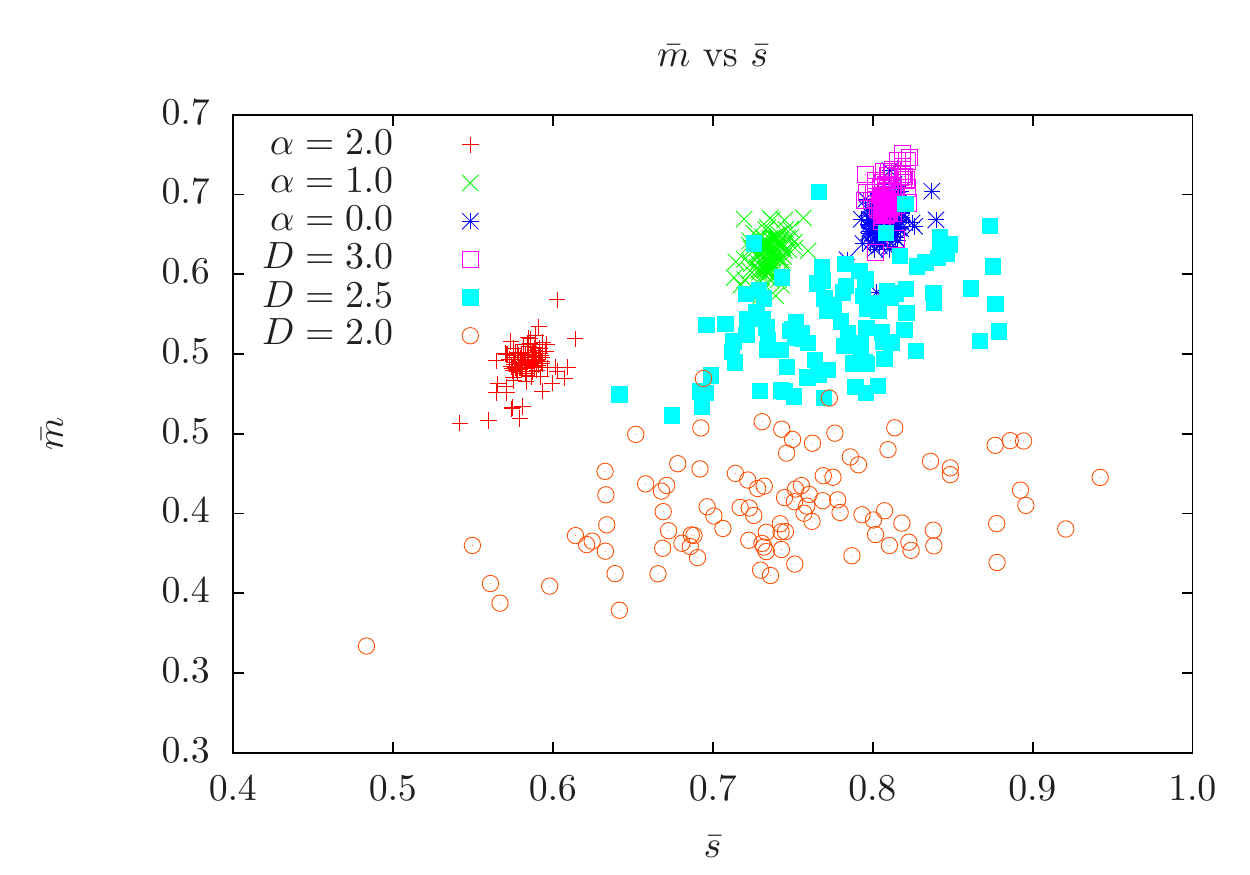}
   \caption{A plot of $\bar{m}$ against $\bar{s}$ for artificial cluster data. Each data set represents one hundred realisations each with one hundred to one thousand points.}
   \label{msbar_clus}
\end{figure}

Tables \ref{r_data} and \ref{f_data} list clustering paramemters for radial and fractal distributions. Column 1 lists the radial density exponent $\alpha$ and fractal dimension $D$. Column 2 lists the size of the images in pixels; rows were $N=\text{``Cluster''}$ relate to artificial star clusters. Columns 3 to 5 give the normalized mean edge length of the MST $\bar{m}$, normalized correlation length $\bar{s}$ and ${\cal Q}=\bar{m}/\bar{s}$. Figures \ref{msbar} and \ref{msbar_clus} show plots of $\bar{m}$ against $\bar{s}$ for $128\times128$ pixel images and artificial clusters respectively. As shown by \citet{qpaper2009} this gives a second means of discriminating $\alpha$ and $D$ by examining which area of the plot specific values of $\bar{m}$ and $\bar{s}$ fall into.

Calculating ${\cal Q}$ does not provide direct information on the structure of a distribution. Instead, $\alpha$ and $D$ are inferred by comparing a measured value of ${\cal Q}$ to that of a distribution of known fractal or radial structure. An example of this is shown in Figure \ref{q_values}, where a fractal dimension or radial density exponent can be estimated by knowing the size of an image $N$ and its measured ${\cal Q}$ value.

The data in Figure \ref{q_values} show how ${\cal Q}$ statistics from grey-scale images vary with image size. As a figure of merit, the closer the data is to ${\mathrm{d}{\cal Q}}/{\mathrm{d}N}=0$, the better the algorithm scales with image size. We observe that for grey-scale data of $2.0\ge\alpha\ge0.0$ and $3.0\ge D\ge2.5$ with image sizes of $4096\ge N\ge65536$ pixels, $N$ has no significant influence on estimating $\alpha$ or $D$.

For images with $N>4000$ and $0.81\ge {\cal Q}\ge0.73$, we find that fractal dimensions in the range $3.0\ge D\ge2.5$ can be estimated from ${\cal Q}$ with an approximate uncertainty of $\sigma_D\sim0.1\,$ irrespective of image size.  For $0.73>{\cal Q}\ge0.54$, estimating $2.5> D\ge2.0$ carries a higher uncertainty of $\sigma_D\sim0.2$ and requires matching ${\cal Q}$ to $N$ as well as $D$. Radial density exponents in the range of $2.0\ge\alpha\ge0.0$ can be estimated when $0.99\ge {\cal Q}\ge0.81\,$. We find that these estimates of $\alpha$ are largely independent of $N$, however, the statistical uncertainties are artifically reduced as, unlike the fractal distributions, the radial mass distributions are completely isotropic.

When comparing ${\cal Q}$ statistics from both artificial clusters and images of artificial clouds, we often find that in distributions of the same $\alpha$ or $D$, clouds have higher ${\cal Q}$ values than clusters. On closer inspection, it can be seen that this arises from higher values of $\bar{m}$ in the cloud data. This can be explained by considering the means by which point distributions are constructed for clusters and clouds. For cluster generation, points are positioned using random numbers (see equation \ref{rcluster}) and therfore are subject to Poisson ``sub-clustering''. When sampling points from images of clouds, whilst the position of the sample area is chosen at random, the sampling area extends over several pixels, thus averaging out some of the small-scale density variation. This goes some way to produce an anti-clustered distribution, which tends to lengthen the MST, thus increasing the value of $\bar{m}$ and ${\cal Q}$. This can be seen in Figure \ref{resplot}, where an increase in $N_\text{pnt}$ reduces the number of pixels from which a point is sampled, reducing anti-clustering and lowering ${\cal Q}$.

It is important to note, that these results relate purely to spherical artifical data. Elongation of star clusters has been shown to have systemattic effect on ${\cal Q}$ \citep{qpaper_elong,SpatialEvolution}, however this can be quantified and corrected for. It is reasonable to assume that this elongation effect also applies for greyscale data and will need to be considered in follow-up work.

\section{Conclusions}
\label{conclusions}

We demonstrate a method and analysis for taking ${\cal Q}$ measurements from grey-scale images of clouds. By decomposing an image into a distribution of points, we are able to apply the pre-existing methods of calculating ${\cal Q}$ and infer information on cloud structure.

Whilst there are systematic differences between ${\cal Q}$ values for clouds and clusters with the same $\alpha$ or $D$, this does not present a problem as the relation between ${\cal Q}$ and $\alpha$ or $D$ can be calibrated independently for both types of data. We also find that grey-scale ${\cal Q}$ values are largely independent of image size for radial density profiles and fractal distributions with $D>2.5\,$. This makes ${\cal Q}$ a powerful tool for studying the structure of molecular clouds alongside that of star clusters.

${\cal Q}$ can also be applied to hydrodynamical simulations. By taking measurements at regular time-steps, the strutural evolution of both gas and sink-particle distribution can be quantified as a function of time.

\section*{Acknowledgments}

Oliver Lomax is a Science and Technology Facilities Council Ph.D student. We thank Stefan Schmeja for the constructive comments and advice relating to this paper.

\bibliographystyle{mn2e}
\bibliography{refs}

\label{lastpage}

\end{document}